\documentclass[preprints,article,accept,oneauthor]{Definitions/mdpi} 
\firstpage{1} 
\pubvolume{xx}
\issuenum{1}
\articlenumber{5}
\pubyear{2020}
\copyrightyear{2020}
\history{Received: 13 September 2020; Accepted: date; Published: date}
\Title{Discrepancy between Power Radiated and the Power Loss Due to Radiation Reaction for an Accelerated Charge}
\Author{Ashok K. Singal\orcidA{0000-0002-8479-7656}}

\address{%
\quad Astronomy and Astrophysics Division, Physical Research Laboratory,
Navrangpura, Ahmedabad - 380 009, India; asingal@prl.res.in
}

\corres{Correspondence: ashokkumar.singal@gmail.com; Tel.: +91-9427-631-833}

\abstract{We examine here the discrepancy between the radiated power, calculated from the Poynting flux at infinity, and the power loss due to radiation reaction for an accelerated charge. It is emphasized that one needs to maintain a clear distinction between the electromagnetic power received by distant observers and the mechanical power loss undergone by the charge. In the literature, both quantities are treated as almost synonymous; the two in general could, however, be quite different. It is shown that in the case of a periodic motion, the two formulations do yield the power loss in a time averaged sense to be the same, even though, the instantaneous rates are quite different. It is demonstrated that the discordance between the two power formulas merely reflects the difference in the power going in self-fields of the charge between the retarded and  present times. In particular, in the case of a uniformly accelerated charge, power going into the self-fields at the present time is equal to the power that was going into the self-fields at the retarded time plus the power going in acceleration fields, usually called radiation. From a study of the fields in regions far off from the time retarded positions of the uniformly accelerated charge, it is shown 
that effectively the fields, including the acceleration fields, remain around the `present' position of the charge which itself is moving toward infinity due to its continuous constant acceleration,  with no other Poynting flow that could be termed as `radiation emitted' by the charge.
}

\keyword{classical electromagnetism; applied classical electromagnetism; radiation by moving charges; radiation or classical fields}

\begin{document}
%%%%%%%%%%%%%%%%%%%%%%%%%%%%%%%%%%%%%%%%%%

%%%%%%%%%%%%%%%%%%%%%%%%%%%%%%%%%%%%%%%%%%

%\shorttitle{Discrepancy between radiated power and radiation reaction losses}
%\shortauthors{Singal}

%-----------------------------------------------
\section{Introduction}
%--------------------------------------------------
In electromagnetic radiation by a point charge, the radiated power  
is proportional to the square of the acceleration, known as Larmor's formula \cite{1,2,25}. On the other hand, the consequent radiation reaction on the charge is directly proportional to the rate of change of the acceleration of the charge \cite{abr05,16,7,20,68b}. The two formulations do not seem to be conversant with each other.
This apparent discordance between the two formulations has survived without a proper, universally acceptable, solution for longer than a century.   
Larmor's formulation is believed to be more rigorous than the radiation-reaction formulation, though there are a large number of arguments, based on energy-momentum conservation laws, that suggest that there is something amiss in Larmor's radiation formula \cite{18,58a}. For instance, the radiation pattern for the slowly moving charge ($v\ll c$) has a $\sin^2\theta$ dependence with respect to the acceleration vector, \cite{1,2,25}, consequently the net momentum carried away by the radiation, averaged over the solid angle, is nil. Therefore, from the momentum conservation law, such a radiating charge cannot suffer any momentum losses. However, due to a finite amount of power going into electromagnetic radiation, as per Larmor's formula, the kinetic energy of the charge must be reducing with time. How could a radiating charge lose kinetic energy without a loss of momentum? Further, an accelerated charge, that may be instantly stationary, has zero kinetic energy at that instant. However, according to Larmor's formula, the charge would be undergoing kinetic energy losses proportional to the square of acceleration, even though its kinetic energy may be zero. 
In order to be still compatible with Larmor's formula, these energy-momentum conservation problems have been circumvented by proposing an acceleration-dependent term, called Schott energy, within electromagnetic fields, that may be lying somewhere in the vicinity of the charge \cite{7,6,44,ER04,56,row12,41}. However, recently, from a critical examination of the electromagnetic fields of a uniformly accelerated charge \cite{58b}, no Schott energy was found anywhere in the near vicinity of the charge, or for that matter, even in the far-off regions.

Here, we critically examine the relation between the two formulations and demonstrate that a mathematical subtlety in the application of Poynting's theorem is being missed when we try to use the energy-momentum conservation laws to compare the two formulas.

We shall, unless otherwise specified, confine ourselves only to non-relativistic motion, as the same set of disparities get carried over to the relativistic case \cite {58a}. Further, we shall assume a one-dimensional motion with acceleration parallel to the velocity and also throughout use the cgs system of units.
%------------------------------------------------------
\section{Two Discrepant Formulations for Radiation Losses from an Accelerated Charge}
The electromagnetic field (${\bf E},{\bf B}$) at a time $t$, of an arbitrarily moving charge $e$, is written as \cite{1,2,25,28},
\begin{eqnarray}
\label{eq:1a}
{\bf E}&=&\left[\frac{e({\bf n}-{\bf v}/c)} {\gamma ^{2}r^{2}(1-{\bf n}\cdot{\bf v}/c)^{3}} + 
\frac{e{\bf n}\times\{({\bf n}-{\bf v}/c)\times \dot{\bf v}\}}{rc^2\:(1-{\bf n}\cdot {\bf v}/c)^{3}}\right]_{\rm t'}\:,
\nonumber\\
{\bf B}&=&{\bf n} \times {\bf E}\:,
\end{eqnarray}
where all quantities in square brackets are to be evaluated at the retarded time $t'=t-r/c$. 

As the acceleration contributes only to the transverse fields, we shall, unless otherwise specified, leave the radial fields aside and consider, henceforth, only the transverse fields.
It is to be emphasized that not only the acceleration fields, even the velocity fields 
have a transverse field component, normal to the radial direction along ${\bf n}$. 

With the help of the vector identity ${\bf v}={\bf n}({\bf v}.{\bf n}) - {\bf n}\times\{{\bf n}\times{\bf v}\}$, transverse components of the electromagnetic field of a charge, having a non-relativistic motion and therefore comprising only linear terms in velocity (${\bf v}$) and acceleration ($\dot{\bf v}$), can be written from Eq.~(\ref{eq:1a}) as
\begin{eqnarray}
\label{eq:1b}
{\bf E}&=&
%e\:\frac{\bf n}{r^{2}}\bigg(1+\frac{2{\bf n} \cdot {\bf v}}{c}\bigg)+
\left[\frac{e{\bf n}\times({\bf n}\times {\bf v})} {cr^{2}} + 
\frac{e{\bf n}\times({\bf n}\times \dot{\bf v})}{c^2r}\right]_{\rm t'}=\left[\frac{e{\bf n}\times({\bf n}\times ({\bf v}+\dot{\bf v}r/c))} {cr^{2}}\right]_{\rm t'}\:,\nonumber\\
{\bf B}&=&
%{\bf n} \times {\bf E}=
\left[-\frac{e{\bf n}\times {\bf v}}{c r^{2}}- \frac{e{\bf n}\times \dot{\bf v}}{c^{2}r}\right]_{\rm t'}=-\left[\frac{e{\bf n}\times ({\bf v}+\dot{\bf v}r/c)}{c r^{2}}\right]_{\rm t'}\:.
\end{eqnarray}

To calculate the radiated electromagnetic power, we make use of the radial component of the Poynting vector \cite{1,2,25}, 
\begin{equation}
\label{eq:8a}
{\bf n} \cdot {\cal S}= \frac{c}{4\pi}{\bf n} \cdot ({\bf E}\times {\bf B})
=\frac{c}{4\pi}({\bf n}\times {\bf E})\cdot {\bf B}=\frac{c}{4\pi}(B)^2.
\end{equation}

Accordingly, 
%using only the acceleration fields (the rightmost term in Eq.~(\ref{eq:1b})),  
one gets for the the radial component of the Poynting vector 
\begin{equation}
\label{eq:8b}
{\bf n} \cdot {\cal S}= \frac{e^2\left[({\bf v}+\dot{\bf v}r/c)^{2}\right]_{\rm t'}}{4\pi r^4 c}\sin^2\theta.
\end{equation}
The $\sin^2\theta$ pattern implies that the rate of momentum being carried in the electromagnetic radiation is zero. 
\begin{equation}
\label{eq:21a1}
\dot{\bf p}_{\rm em}=0\:.
\end{equation}
However, the net Poynting flow through a spherical surface, $\Sigma$ of radius $r$, around the charge, for a large $r$, is
\begin{equation}
\label{eq:21a}
{\cal P}_{\rm em}= \int_{\rm r\rightarrow\infty}{{\rm d}\Sigma}\:({\bf n} \cdot {\cal S})
=\frac{e^2}{2 c}\int_{0}^{\pi} {\rm d}\theta\: \sin^3\theta\:\left.\frac{\left[({\bf v}+\dot{\bf v}r/c)^{2}\right]_{\rm t'}}{r^2 }\right|_{\rm r\rightarrow\infty}
=\frac{2e^{2}}{3c^{3}} \left[\dot{\bf v}^{2}\right]_{\rm t'}.
\end{equation}
This is Larmor's famous result for the electromagnetic power radiated from an accelerated charged particle \cite{1,2,25}.
Since the contribution of velocity fields ($\propto 1/r^{2}$), for a large enough $r$, seems negligible, with the Poynting flow due to the acceleration fields being independent of $r$ (Eq.~(\ref{eq:21a})), the common perception is that in {all cases}, the acceleration fields ($\propto 1/r$) alone represent radiation from a charge.
% being proportional to the square of its acceleration; the power loss occurring out of the kinetic energy of the charge. 

Presumably, using the energy-momentum conservation laws \cite{ha95,51}, we can compute the mechanical energy-momentum losses of the radiating charged particle.
% as a consequence of electromagnetic radiation emitted, we get inconsistent results. 
For instance, the momentum of the charge would not change due to radiation damping,
\begin{equation}
\label{eq:21b2}
{\bf F}=-\dot{\bf p}_{\rm em}=0\:,
\end{equation}
while the kinetic  energy, $\cal T$, of the charged particle should change due to radiation losses at a rate 
\begin{equation}
\label{eq:11a} 
\frac{{\rm d}{\cal T}}{{\rm d}t}=-{\cal P}_{\rm em}.
\end{equation}  
Now, Eqs.~(\ref{eq:21b2}) and (\ref{eq:11a}) do not seem mutually consistent since the charged particle cannot lose kinetic energy without losing momentum. In fact, some problem is inherently present in Eq.~(\ref{eq:11a}) itself, as in the rest frame of the charge, the energy loss rate is finite ($\propto \dot{\bf v}^2$) even when the charged particle has no kinetic energy (${\bf v}=0)$ to lose. It may be pointed out here that such a power loss into radiation can happen, without any change in the kinetic energy of the emitting charge, only if there were a loss of internal (rest mass!) energy, without an accompanying loss of momentum \cite{mo94}. However, we do not contemplate a radiating charged particle to be converting its rest mass energy into electromagnetic radiation; after all, a radiating electron still remains an electron at the end of the emission of radiation. 

Somewhere something is amiss!
\subsection{An Inappropriate Usage of the Poynting Theorem}
Actually, in the above formulation, which is the standard text-book approach, one is equating   the Poynting flux at time $t$ to the mechanical power loss of the 
charge at a retarded time $t-r/c$, purportedly using Poynting's theorem of energy conservation. 
However, there is a fallacy in this particular step as Poynting's theorem does 
not directly relate the Poynting flux through a closed surface at a time $t$ to power losses by the enclosed charge at a retarded time $t-r/c$. Since the electromagnetic fields at $r$ at time $t$ do get determined by the charge motion at the retarded time $t'=t-r/c$, one may intuitively be tempted to equate the electromagnetic power represented by the Poynting flux at $r$ at time $t$ to the mechanical power {loss} of the charge at the retarded time $t'=t-r/c$. Our common-sense notion of causality may, however, be in conflict with the strict mathematical definition of Poynting's theorem and the ensuing application of energy-momentum conservation laws to  electromechanical systems could sometimes lead us astray.
It is such an oversight in this case that has mostly been the cause of confusion in this long drawn out  controversy. 
In Poynting's theorem, all quantities are to be calculated, strictly for the {same instant of time} \cite{1,2,25}. 

Applying the Poynting's theorem in an appropriate manner,  one obtains the 
instantaneous mechanical power loss of the charge, in terms of the real time values of the charge motion \cite{68a}, as 
\begin{equation}
\label{eq:21b}
{\cal P}_{\rm me} =-\frac{2e^{2}}{3c^{3}}\ddot{\bf v}\cdot{\bf v}\:,
\end{equation}
with all quantities evaluated for the same common instant, say, $t$.

Now, one needs to maintain a clear distinction between the electromagnetic power received by  distant observers (Eq.~(\ref{eq:21a})) and the instantaneous mechanical power loss of the charge (Eq.~(\ref{eq:21b})). In the literature both power rates are treated as almost synonymous. However, as we can see, the two are not necessarily the same (cf. Eqs.~(\ref{eq:21a}) and (\ref{eq:21b})). 

The difference in the two power formulas is 
\begin{equation}
\label{eq:9.1}
{\cal P}_{\rm me}-{\cal P}_{\rm em}=-\frac{2e^{2}}{3c^{3}}\ddot{\bf v}\cdot{\bf v} -\frac{2e^{2}}{3c^{3}}\left[\dot{\bf v}\cdot\dot{\bf v}\right]_{\rm t'}
=-\frac{2e^{2}}{3c^{3}}\frac{{\rm d}(\dot{\bf v}\cdot{\bf v})}{{\rm d}t}\:.
\end{equation}
The rightmost term in Eq.~(\ref{eq:9.1}) is the total time derivative of a term, known as  Schott energy, believed to be an acceleration-dependent energy, $-{2e^{2}(\dot{\bf v}\cdot{\bf v})}/{3c^{3}}$, in the electromagnetic fields, lying in the near vicinity of the charge \cite{7,6,44,ER04,56,row12,41}. This elusive, century-old term does not seem to have been encountered elsewhere in physics. 
As we shall demonstrate in Section 2.3, the Schott term is not any real electromagnetic energy in the fields and makes an appearance in the above equation merely because the power going in  the self-field of an accelerated charge between``real'' and ``retarded'' times is different. 

In the same way, exploiting Maxwell's stress tensor, from the momentum conservation theorem we  can arrive at the expression for the rate of change of momentum of the charge \cite{68b}, written as 
\begin{equation}
\label{eq:3a}
\dot{\bf p}_{\rm me}= \frac{2e^{2}}{3c^{3}}\ddot{\bf v}\:.
\end{equation}
The result in  Eq.~(\ref{eq:3a}), is known as the Abraham--Lorentz radiation reaction formula, and was earlier derived from a detailed computation of the self-force due to a rate of change of  acceleration ($\ddot{\bf v}$) of the instantly stationary (${\bf v}=0$) charge, in a quite involved manner \cite{1,2,7,abr05,16,20}.
%----------------------------------------------------------
\subsection{Applicability of Larmor's Formula to Compute Radiative Power Losses in Case of a Periodic Motion}
%-----------------------------------------------
Does the discrepancy in two formulations imply that Larmor's formula cannot be applied for computing mechanical power losses for a radiating charge? 

In the case of a periodic motion of period $T$, there is no difference in the radiated power integrated or averaged between $t$ to $t+T$ and  $t'$ to $t'+T$, therefore  Larmor's formula, does yield a correct average power loss by the charge for a periodic case. 

Let us write the motion of a harmonically oscillating charge (like in a radio antenna) as  
\begin{equation}
\label{eq:10.1}
%{\bf x}={\bf x}_1 \cos(\omega t+\psi_1)+{\bf x}_2 \sin(\omega t+\psi_2)\;.
{\bf x}={\bf x}_{\rm o} \sin(\omega t+\psi)\;.
\end{equation}
Then
\begin{equation}
\label{eq:10a}
%{\bf v}=\dot{{\bf x}}=\omega[-{\bf x}_1 \sin(\omega t+\psi_1)+{\bf x}_2 \cos(\omega t+\psi_2)]
{\bf v}=\dot{{\bf x}}=\omega{\bf x}_{\rm o} \cos(\omega t+\psi)\;,
\end{equation}
\begin{equation}
\label{eq:10b}
%\dot{{\bf v}}=\ddot{{\bf x}}=-\omega^2 [{\bf x}_1 \cos(\omega t+\psi_1)+{\bf x}_2 \sin(\omega t+\psi_2)=-\omega^2 {\bf x}\;,
\dot{{\bf v}}=\ddot{{\bf x}}=-\omega^2 {\bf x}_{\rm o} \sin(\omega t+\psi)=-\omega^2 {\bf x}\;,
\end{equation}
\begin{equation}
\label{eq:10c}
%\ddot{{\bf v}}=-\omega^3 [-{\bf x}_1 \sin(\omega t+\psi_1)+{\bf x}_2 \cos(\omega t+\psi_2)]=-\omega^2 {\bf v}\;.
\ddot{{\bf v}}=-\omega^3 {\bf x}_{\rm o} \cos(\omega t+\psi)=-\omega^2 {\bf v}\;.
\end{equation}

For such a motion, one gets from Larmor's formula the radiative power $\propto \dot{\bf v}^2=\omega^4 {\bf x}^2_{\rm o} \sin^2(\omega t+\psi)$ while the radiation reaction formula yields a  power loss $\propto -\ddot{\bf v}\cdot{\bf v}= \omega^4 {\bf x}^2_{\rm o} \cos^2(\omega t+\psi)$. Both   
expressions yield the same result when averaged over a full cycle, however, the instantaneous rates are very different. 
It means the power spectrum, which gives average power at each frequency, would be the same in most cases, where the actual motion of the charge could be Fourier analysed.
Of course in a non-periodic case like that of a uniformly accelerated charge, where a Fourier analysis is not possible, the two formulas could yield discordant answers. 
%-----------------------------------------------------
\subsection{Discrepancy in Two Power Formulas is Due to the Difference in Power Going in Self-Fields at `Real' and Retarded Times}
%----------------------------------------------------------
\begin{figure}[t]
\begin{center}
\includegraphics[width=7cm]{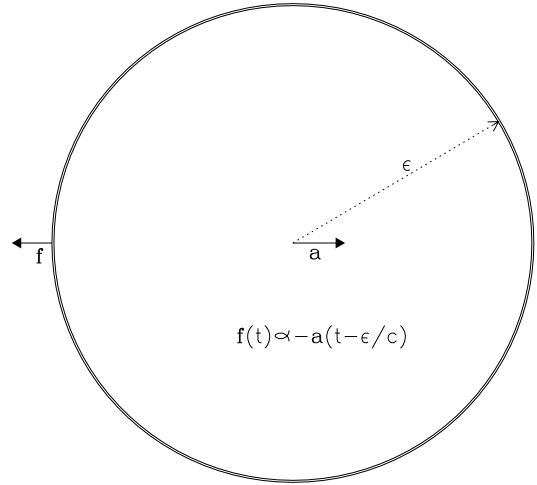}
\caption{The self-force ${\bf f}$ on a charged spherical shell of a small radius $\epsilon$, moving non-relativistically with an acceleration ${\bf a}=\dot{\bf v}$. The net self-force on the charged shell at any instant is proportional to the acceleration it had at a time  interval $\epsilon /c$ earlier \cite{68}. This implies that for a uniform acceleration $a$, the self-force on the charge would only be an `inertial' force  $-m_{\rm el} a$, where $m_{\rm el}=2e^{2}/3\epsilon c^{2}$ is the electromagnetic mass of the charge \cite{FE64}, without any radiation reaction, whatsoever, consistent with the fact that there is no radiation emitted in this case.}
\end{center}
\end{figure}
%----------------------------------------------------------
In order to understand the genesis of the difference between Eqs.~(\ref{eq:21a}) and (\ref{eq:21b}), which respectively are at the retarded and real times, we compute the rate of work done  by the self-force of an accelerated charge. We consider the charge to be in the shape of a spherical shell of a sufficiently small radius $\epsilon$, though, as we shall see, the final results sought by us will be independent of the value of $\epsilon$. We compute the force on each infinitesimal element of the charged shell due to the time-retarded fields from the remainder parts of the  shell and perform a double integration over the shell to get the total self-force on the charge \cite{1}. 
The net self-force at a time $t$ on the accelerated charged spherical shell of radius $\epsilon$ turns out to be proportional to the acceleration it had at a retarded time $t'=t-\epsilon /c$ (Fig.~1) \cite{68}.
\begin{equation}
\label{eq:3d1}
{\bf f}_{\rm t}=-\frac{2e^{2}}{3\epsilon c^{2}}\dot{\bf v}_{\rm t'}\:,
\end{equation}
where $\dot{\bf v}_{\rm t'}$ is the acceleration of the charge at the retarded time $t'$.
Then, for the charge moving with velocity $v_{\rm t}$ at time $t$, the work being done against self-force of the charge is  
\begin{equation}
\label{eq:6}
{\cal P}_{\rm t}=-{\bf f}_{\rm t} \cdot {\bf v}_{\rm t}=\frac{2e^{2}}{3\epsilon c^{2}}\dot{\bf v}_{{\rm t'}} \cdot {\bf v}_{\rm t}\:.
\end{equation}
Because work is done against the self-force, this is the rate at which energy is being put into the fields of the charge.

By expressing the acceleration at time $t'$ in terms of its real-time value at $t$, to a first order, we have 
\begin{equation}
\label{eq:22}
\dot{\bf v}_{{\rm t}'} =\dot{\bf v}_{\rm t}-{\ddot{\bf v}}\epsilon/c. 
\end{equation}
Then from Eq.~(\ref{eq:3d1}) for the self-force, in terms of real-time values, we can write 
\begin{equation}
\label{eq:3a1}
{\bf f}_{\rm t}= -\frac{2e^{2}}{3\epsilon c^{2}}\dot{\bf v}_{\rm t}+\frac{2e^{2}}{3c^{3}}\ddot{\bf v}_{\rm t}\:,
\end{equation}
where the last term is the well-known Abraham--Lorentz radiation reaction (Eq.~(\ref{eq:3a})). 

From this we get the formula for power going into the fields (Eq.~(\ref{eq:6})), but now expressed in terms of the real time values, as
\begin{equation}
\label{eq:6a}
{\cal P}_{\rm t}=\frac{2e^{2}}{3\epsilon c^{2}}(\dot{\bf v}\cdot{\bf v})_{\rm t}
-\frac{2e^{2}}{3c^{3}}(\ddot{\bf v}\cdot{\bf v})_{\rm t}\:.
\end{equation}
The first term on the right hand side is the rate of change of self-field energy of the accelerated charge {\em at the real time} $t$, and the second term is power loss due to radiation reaction (Eq.~(\ref{eq:21b})), again evaluated at $t$.

On the other hand, if we express the velocity itself in terms of its 
value at the retarded time ${t'}$, to a first order in $\epsilon/c$, we have  
\begin{equation}
\label{eq:21.1}
{\bf v}_t={\bf v}_{{\rm t}'}+\dot{\bf v}_{{\rm t}'}\:{\epsilon}/{c}.
\end{equation}
Substituting in (Eq.~(\ref{eq:6})), we get  
\begin{equation}
\label{eq:9}
{\cal P}_t=\frac{2e^{2}}{3\epsilon c^{2}}[\dot{\bf v}\cdot
{\bf v}]_{{\rm t}'}+\frac{2e^{2}}{3c^{3}}[\dot{\bf v}\cdot\dot{\bf v}]_{{\rm t}'} \:.
\end{equation}
The first term on the right hand side shows the rate of change of self-field energy of the accelerated charge 
%due to its changing momentum 
{\em at the retarded time} $t'$, and the second term is Larmor's formula  (Eq.~(\ref{eq:21a})), again evaluated at $t'$.

From Eqs.~(\ref{eq:6a}) and (\ref{eq:9}), we get
\begin{equation}
\label{eq:9.1a}
%{\cal P}_{\rm me}-{\cal P}_{\rm em}=
-\frac{2e^{2}}{3c^{3}}(\ddot{\bf v}\cdot{\bf v})_t -\frac{2e^{2}}{3c^{3}}[\dot{\bf v}\cdot\dot{\bf v}]_{{\rm t}'}
=\frac{2e^{2}}{3\epsilon c^{2}}[\dot{\bf v}\cdot {\bf v}]_{t'}-\frac{2e^{2}}{3\epsilon c^{2}}(\dot{\bf v}\cdot{\bf v})_{{\rm t}}
%=-\frac{2e^{2}}{3c^{3}}\frac{{\rm d}(\dot{\bf v}\cdot{\bf v})}{{\rm d}t}\:.
\end{equation}
It shows that the difference in the two power formulas, Eqs.~(\ref{eq:21b}) and (\ref{eq:21a}), which respectively are at real and retarded times, is merely the difference in the power going  in self-fields of the charge between retarded and  present times.

Now, we can write the right hand side of Eq.~(\ref{eq:9.1a}) as
\begin{eqnarray}
%\nonumber
\label{eq:9.2}
\frac{2e^{2}}{3\epsilon c^{2}}[\dot{\bf v}\cdot {\bf v}]_{\rm t'}-\frac{2e^{2}}{3\epsilon c^{2}}(\dot{\bf v}\cdot{\bf v})_{{\rm t}}
=-\frac{2e^{2}}{3\epsilon c^{2}}\frac{{\rm d}(\dot{\bf v}\cdot{\bf v})}{{\rm d}t}\frac{\epsilon}{c}
=-\frac{2e^{2}}{3c^{3}}\frac{{\rm d}(\dot{\bf v}\cdot{\bf v})}{{\rm d}t}\:,
\end{eqnarray}
a result independent of $\epsilon$. This demonstrates that the elusive Schott term is not some actual hidden energy in the near fields but shows up in Eq.~(\ref{eq:9.1}) merely due to the different rates of energy change in the self-fields between retarded and present times of an accelerated charge. This is consistent with the findings from a critical examination of the electromagnetic fields of a uniformly accelerated charge \cite{58b}, where, contrary to the suggestions in the literature  \cite{6,44,ER04,56,row12,41}, no Schott energy term was found anywhere in the near vicinity of the charge, or for that matter, even in far-off regions.

%-----------------------------------------------
\section{A Uniformly Accelerated Charge}
In the derivation of Larmor's formula (Eq.~(\ref{eq:21a1})), which is a standard text-book material \cite{1,2,25}, it is assumed that any contribution of velocity fields could be neglected. 
This assumption holds true in almost all cases, except in that of a uniformly accelerated charge, where the velocity may change monotonically with time \cite{18}. 

Actually, in the case of a uniform acceleration, in the expressions for the fields (Eq.~(\ref{eq:1b})), the {retarded} velocity of the charge would be related to the {present value} of velocity, ${\bf v}_{\rm o} = [{\bf v} +\dot{\bf v} r/c]_{\rm t'}$. Then the transverse components of the electromagnetic fields become 
%\begin{align}
\begin{eqnarray}
\label{eq:1a11}
{\bf E}&=&
%e\:\frac{\bf n}{r^{2}}\bigg(1+\frac{2{\bf n} \cdot {\bf v}}{c}\bigg)+
\frac{e{\bf n}\times({\bf n}\times {\bf v}_{\rm o})} {cr^{2}} 
%+ \frac{e{\bf n}\times({\bf n}\times \dot{\bf v})}{c^2r}
\:,\nonumber\\
{\bf B}&=&\frac{-e{\bf n}\times {\bf v}_{\rm o}} {cr^{2}}.
\end{eqnarray}
%\end{align}
Thus we see that what all the acceleration fields do in this case is to make the instantaneous transverse fields {everywhere} directly proportional to the instantaneous present velocity $v_{\rm o}$ of the accelerated charge. 

%-------------------------------------------------------------------
\subsection{The Contribution of Acceleration Fields to the Energy-Momentum of Self-Field}
Since the self-field energy of a charge moving with a uniform velocity depends upon the magnitude of the velocity (see, e.g.,~\cite{18}), when a charge is accelerated, its self-field energy must change too, depending upon the change in velocity. 
It is therefore imperative that the acceleration fields provide for the changes taking place in the energy in self-fields.
% as the change in self-field energy cannot come from the velocity fields 
%alone  which contains no information about the change that might take place in the velocity of the charge.
%which are  attached to the charge. 
As the acceleration, $\dot{\bf v}$, changes the velocity of the charge to say, ${\bf v}_{\rm o} ={\bf v} + \dot{\bf v} r/c$, the acceleration fields ($\propto\dot{\bf v}/r$) ensure that the transverse fields accordingly remain `updated' ($\propto {\bf v}_{\rm o}/r^2$), to remain synchronized with the `present' value of the velocity of the charge, and the energy in self-fields is always equal to that required because of the `present' velocity of the accelerated charge. The conventional wisdom, on the other hand, is that the acceleration fields, exclusively and wholly, represent power irreversibly lost as radiation, given by Larmor's formula. Thus there may be something amiss in the standard picture where one does not even consider that the Poynting flux from the acceleration fields might be contributing toward the changing self-field energy of the accelerating charge. After all a stationary charge has zero self-field energy in transverse fields, and the growth in the self-field energy as the charge picks up speed due to acceleration, could have come only from the acceleration fields.
The radiation actually would only be that part of the Poynting flux which is over and above the value determined by the change occurring in the instantaneous  velocity of the charge. 

Employing the formula for the electromagnetic field energy  
\begin{eqnarray}
\label{eq:89.4k}
{\cal E}= \int_{\rm V} \frac{E^{2}+B^{2}}{8\pi}\:{\rm d}V \;,
\end{eqnarray}
it is possible to compute the electromagnetic field energy, not only for a charge moving with a uniform velocity, but even in the case of a charge moving with a uniform acceleration \cite{18}. For instance, the transverse field energy of the uniformly accelerated charge, in a shell of volume $4\pi r^{2} {\rm d}r$, enclosed between spheres $\Sigma$ and $\Sigma_1$ of radii $r$ and $r+{\rm d}r$, is
\begin{eqnarray}
\label{eq:89.4m}
{\rm d}{\cal E} =\frac{e^{2}}{2}\left(\frac{4 v_{\rm o}^{2}}{3c^2}\right)\,
\frac{{\rm d}r}{r^{2}} \;.
\end{eqnarray}
We can integrate over $r$ to get the total energy in the transverse fields outside a sphere of radius $\epsilon$ as,
\begin{equation}
\label{eq:89.4p}
{\cal E} =\frac{2e^{2}}{3c^2\epsilon}v_{\rm o}^{2}\;.
\end{equation}
Since the integral diverges for $r\!\rightarrow \!0$, we restricted the lower limit of $r$
to a small $\epsilon$, which may represent the radius of the charged particle. 

One can also calculate the energy in fields of a charge moving with a uniform velocity $v_{\rm o}$ and exactly the same amount of field energy is found around the charge. 
Thus it is clear that the acceleration fields in the case of a  uniformly accelerated charge add just sufficient energy in the self-fields so as to make the total field energy equal to that required because of the `present' velocity of the accelerating charge. This is true even in the case of the charges moving with relativistic velocities \cite{18}.

%If we consider a charge with uniform acceleration ${a}=\dot{v}$, instantly stationary (i.e., ${v}=0$) at time $t=0$, then the energy in its transverse fields is zero. After a time $t$, when it is moving with a velocity $v_{\rm o}= at=ar/c$, due to its `present' velocity $v_{\rm o}$, an energy in its transverse self-fields in a spherical shell between $r$ and $r+dr$ would be,
%\begin{equation}
%\label{eq:89.4n}
%d{\cal E} =\frac{2e^{2}}{3c^{2}}(v_{\rm o}/c)^{2}\frac{dr}{r^2}
%=\frac{2e^{2}}{3c^{3}}a^{2}{dt}\;,     
%\end{equation}
%the last term on the right hand side exactly accounting for the 
%commonly--believed energy `radiated away' during the time interval $dt=dr/c$, which actually goes into building the self-fields of the uniform accelerated charge.
%%----------------------------------------------

That the Poynting flux in the acceleration fields feeds the self-field energy in the case of a uniformly accelerated charge, is further seen from a comparison of the self energy changes between the real and the retarded times. Since in the case of a uniformly accelerated charge $\ddot{\bf v}=0$, then from Eq.~(\ref{eq:9.1a}), we get
\begin{equation}
\label{eq:9.1b}
\frac{2e^{2}}{3\epsilon c^{2}}(\dot{\bf v}\cdot{\bf v})_{\rm t}=\frac{2e^{2}}{3\epsilon c^{2}}[\dot{\bf v}\cdot {\bf v}]_{{\rm t}'}+\frac{2e^{2}}{3c^{3}}[\dot{\bf v}\cdot\dot{\bf v}]_{{\rm t}'}\:.
\end{equation}
From Eq.~(\ref{eq:9.1b}), it is obvious that in the case of a uniformly accelerated charge, power going into the self-fields at the present time $t$ is equal to the power that was going into the self-fields at the retarded time $t'$ plus the power going in acceleration fields, usually called Larmor's formula for radiative losses. Instead of any losses being suffered by the charge, the energy in its self-fields is actually being constantly augmented by the acceleration fields.  
There is no other power term in the formulation that could be called {radiation emitted} by the uniformly accelerated charge.

We can compute the net momentum as well, in the self-fields of a uniformly accelerated charge, 
from the volume integral 
\begin{eqnarray}
\label{eq:89.4q}
{\bf p}= \int_V \frac{\bf E\times B}{4\pi c}\:{\rm d}V\:.
\end{eqnarray}
Due to the azimuth symmetry about the direction of motion, the transverse component of the electric field (Eq.~(\ref{eq:1a11})) makes a nil contribution to the momentum, when integrated over the solid angle. However, the radial component, $e{\bf n}/r^2$, does make a net finite contribution, which would be along the direction of motion. Accordingly, we get
\begin{eqnarray}
\label{eq:89.4r}
{\bf p}= \frac{e^2{\bf v}_{\rm o} }{2\epsilon c^2}\int_{0}^{\pi} {\rm d}\theta\: \sin^3\theta
=\frac{2e^{2}}{3\epsilon c^{2}} {\bf v}_{\rm o}= m_{\rm el}{\bf v}_{\rm o} \;,
\end{eqnarray}
where $m_{\rm el}=2e^{2}/3\epsilon c^{2}$ is the electromagnetic mass of the charge \cite{FE64}.
Thus we see that as the charge velocity changes to ${\bf v}_{\rm o}$ due to the acceleration,  the acceleration fields contribute to the self-fields of the charge, so that the field momentum becomes $m_{\rm el}{\bf v}_{\rm o}$, in accordance with the instantaneous velocity ${\bf v}_{\rm o}$.

Thus both the energy and momentum in the self-fields of the uniformly accelerated charge are getting constantly updated by its acceleration fields in accordance with its `present' velocity at any instant.
%-----------------------------------------------------------------------------------
\subsection{Poynting Flux in the Case of a Uniformly Accelerated Charge}
In the derivation of  Larmor's formula (Eq.~(\ref{eq:21a})), one assumed that the velocity fields would {always} make a negligible contribution to the Poynting flow, for large $r$. 
However, in the case of a uniformly accelerated charge, the contribution of velocity fields could match that of the acceleration fields, for all $r$. 
From Eq.~(\ref{eq:1a11}), we find the Poynting flux to be 
\begin{equation}
\label{eq:11c1}
{\cal P}= \frac{2e^{2} {\bf v}_{\rm o}^{2}}{3r^2c}.
\end{equation}
The power passing through the spherical surface in the case of a uniformly accelerated charge is $\propto  v_{\rm o}^{2}/r^2$.

A similar transverse component of the electromagnetic field (Eq.~(\ref{eq:1a11})) is also seen in the case of a charge moving with a uniform velocity ${{\bf v}}_{\rm o}$, equal to the ``present'' velocity of the accelerated charge. 
Therefore, a Poynting flux exactly similar to Eq.~(\ref{eq:11c1}) is also present in the case of a uniformly moving charge, where we know there are no radiation losses and the Poynting flow through a surface around time-retarded position of the charge is merely due to the ``convective'' flow of fields, along with the moving charge. However, with respect to the `present' position of a charge, there is no radial Poynting flux in this case. Taking a cue from this, even for a uniformly accelerated charge, one should examine the Poynting flux vis-\`a-vis the `present' position of the accelerated charge, to find out if there indeed is some radiation taking place.
%In regions outside the spherical surface of radius $r=ct$, in the case of a uniformly accelerated charge, it accounts for the change occurring in the Coulomb field energy of the charge, because of the ever changing velocity of the charge.
%The charge after all, now moving with $v_{\rm o}$, with a Poynting flux $\propto v_{\rm o}^2$ was initially had $v_{\rm o}=0$, with a nil Poynting flux, and this increase in the Poynting flux has come only due to the acceleration fields.
As the energy in the self-fields must be "co-moving" with the charge, 
(otherwise the self-fields would lag behind, and no longer remain about the charge to qualify as its self-fields), and there should accordingly be a Poynting flow. Therefore not all of the Poynting flow may constitute radiation. The radiated power would be the part of the Poynting flow that is detached from the charge \cite{25}, i.e., it should be over and above the energy changes in the self-fields of the charge, as determined from the changing velocity of the charge. As we saw from the energy-momentum in the fields in Section 3.1, there is no such excess energy in fields to be termed as radiation in the case of a uniformly accelerated charge.
%
%It is clear that in the case of a uniformly accelerated charge, in the Poynting flux expression (Eq.~(\ref{eq:11c1})), there is no term proportional to $\dot{\bf v}^{2}$, 
%independent of $r$, which is commonly called the radiated power. Instead, the Poynting flux in Eq.~(\ref{eq:11c1}) is exactly that from a charge moving with a uniform velocity ${{\bf v}}_{\rm o}$.
%The Poynting flux in Eq.~(\ref{eq:11c1}) at time $t$ due to the "present" velocity of the accelerating charge 
%can be related to the erstwhile called radiated power (Eq.~(\ref{eq:21a})) by substitution of ${\bf v}_{\rm o} =\dot{\bf v}r/c$ to get,  
%\begin{equation}
%\label{eq:1r1}
%\frac{2e^2{\bf v}_{\rm o}^2}{3r^{2}c}=\frac{2e^2\dot{\bf v}^2r^{2}}{3r^{2}c^3}=\frac{2e^2\dot{\bf v}^2}{3c^3} \;.
%\end{equation}
%Thus we see that in the case of a constant acceleration, the hitherto termed radiation losses (\`{a} la Larmor's formula) 
%are nothing but the Poynting flow due to the movement of self-fields along with the
%charge, because to its "present" velocity ${\bf v}_{\rm o}$.    

It is evident from Eq.~(\ref{eq:1a11}) that 
the transverse component of electromagnetic field, at least in the instantaneous rest frame (${{\bf v}}_{\rm o}=0$) of a uniformly accelerated charge, is nil. This happens due to a systematic cancellation of acceleration fields by the transverse component of velocity fields, in the instantaneous rest frame, both for the electric and magnetic fields, {at all distances}. 
That the magnetic field is zero everywhere in this case was first pointed out by Pauli \cite{33}, using Born's solutions \cite{32}, who inferred from it that no wave zone would be formed and hence there is no radiation from a uniformly accelerated charge.

% Consequently there is a nil Poynting flux through any surface around such a  charge (Eq.~(\ref{eq:11c1})), and a nil radiation reaction on the charged particle, which is consistent with Eqs.~(\ref{eq:21b}) and (\ref{eq:3a}) as there is no rate of change of acceleration ($\ddot{v}=0)$ in this case. 
%--------------------------------------------------------------------------
\subsubsection{A definition of radiation at infinity incompatible with Green's theorem}
It has been claimed that Pauli's statement, that contradicts Larmor's formula, is invalid on the grounds that a limit to large $r$ at a fixed time, say, $t = 0$, is implied therein \cite{5,5a}. It has been asserted that the radiation should instead be defined by the total rate of energy emitted by the charge at the retarded time $t'$, and is to be calculated by integrating over the surface of the light sphere in the limit of infinite $r=c(t-t')$ for a {fixed emission time} $t'$, with both $t\rightarrow\infty$ and $r\rightarrow\infty$ \cite{5,5a}. The two limiting procedures, one with $t$ fixed and the other with $t'$ fixed, do not yield the same result and from that it has been concluded that Pauli's observation that  $B=0$ everywhere at some fixed time $t$ is a mere curiosity that may be of some interest but 
%had nothing to do with the question of radiation \cite{5,5a}.
does not imply an absence of radiation \cite{5,5a}.

If we carefully examine the reason why a {fixed emission time} $t'$ is being chosen for  defining `radiation' \cite{5,5a}, we can see that this choice makes the contribution to the Poynting flow, from the velocity fields at $t'$, for a large enough $r$, negligible. However, for a uniformly accelerated charge, one cannot ignore the contribution of the velocity fields to the Poynting flow, as $v(t')\propto -\dot{v}r/c$. Moreover, in this case, there is something unusual happening about the fields at large $r$ vis-\`a-vis the charge location at large $t$, which we shall discuss in Section 3.3.

Actually in Green's retarded solution, the scalar potential $\phi$ at a field point $\bf x$, for instance, is determined at time $t$ from the volume integral  \cite{1}
\begin{eqnarray}
\label{eq:90.1}
%\phi(\bf X,t)&=&\int[\frac{\left\rho(\bf X')\right]}{|{\bf {x-x'}}|}_{t_{\rm o}}\nonumber\\
\phi(\bf x,t)&=&\int\frac{\left[\rho(\bf x')\right]_{t'}}{r}{\rm d}^3 x'\;.
\end{eqnarray}
Here the charge density $\rho(\bf x')$ at $\bf x'$, enclosed within square brackets, and at a distance $r=|{\bf {x-x'}}|$ from the field point, is to be determined at the retarded time $t'=t-r/c$. A similar expression is there for the vector potential as well. 

Thus here ${\bf x}$ and $t$ are specified first and the volume integral of $\rho/r$ at the corresponding retarded times is then computed. Pauli's argument is consistent with this procedure. In fact, the radiation defined by first fixing the emission time, $t'$ \cite{5,5a}, strictly speaking, may not be in tune with Green's retarded time solution, and could sometime lead to wrong conclusions, especially in the limit $r\rightarrow\infty$. 

It may be pointed out that for a ``point'' charge e moving with velocity $\bf v$, first fixing the point charge position ${\bf x'}$ at the retarded-time $t'$, to determine the potential this way, yields $\phi=e/r$, while the more correct approach of first fixing the field point $\bf x$ at time $t$, leads to $\phi=e/[r(1-{\bf n}\cdot{\bf v}/c)]$, the correct expression for the potential \cite{FE64}.
%-----------------------------------------------------------------------------------
\subsection{Far Fields and the Relative Location of the Uniformly Accelerated Charge}
Conclusions about radiation from a uniformly accelerated charge, contrary to ours, seem to have  been drawn previously. This was because, firstly only the acceleration fields were being considered, an approach which though might be valid in vast majority of cases of radiation from an accelerated charge, but is not valid in the case of a uniformly accelerated charge. The reason being that in the latter case the velocity at the retarded time being $v\propto \dot{v}t=\dot{v}r/c$, the velocity fields, $v/r^2\propto \dot{v}r/cr^2$ become comparable to the acceleration fields, $\propto \dot{v}/rc$, for all $r$. Secondly, almost no attention has generally been paid to the `present' location of the charge vis-\`a-vis the fields that move to   $r=ct$. As we will show, during the intervening time interval $t=r/c$, the charge is almost keeping in step with the fields, being only a finite distance $\propto 1/\gamma$ behind for all $t$, with $\gamma$ ever increasing due to the uniform acceleration.
As such, the fields remain appreciable along the direction of motion only in a small, finite region $\propto 1/\gamma$ about the `present' position of the charge, very similar to the  uniform velocity case where electric field is ever appreciable only near the `present' position of the charge, in a region whose extent falls as $1/\gamma$ and where the field strength is mostly along the direction normal to the direction of motion (see, e.g., \cite{88}). In the literature, almost no attention has been paid to the charge position relative to the light-front of the supposed to be radiation fields or vice-versa, in the case of a uniformly accelerated charge.

Since we want to examine far fields at large $r$, this would also imply large values of $t=r/c$. Now, a uniform acceleration for a long duration could make the motion of the charge relativistic, accordingly, in this Section, we shall no longer assume the motion to be non-relativistic.
%----------------------------------------------------------
\begin{figure*}[t]
\begin{center}
\includegraphics[width=15cm]{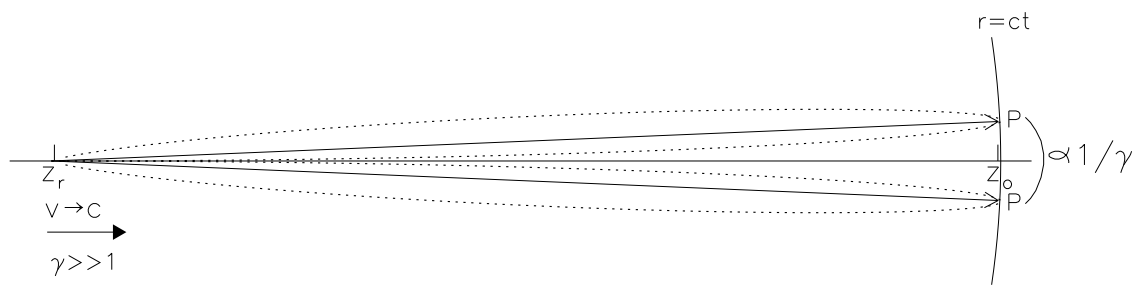}
\caption{Angular distribution of the electric field strength with respect to the time-retarded position $z_{\rm r}$ of the uniformly accelerated charge, moving along the $z$-axis with  velocity $v\rightarrow c$ and the corresponding Lorentz factor $\gamma\gg 1$. Due to the relativistic beaming, the field strength is mostly appreciable only within a cone of angle $\theta \sim 1/\gamma$ about the direction of motion. When at time $t$, the fields from the retarded position $z_{\rm r}$ are at the spherical light-front of radius $r=ct$, the charge meanwhile has moved to $z_{\rm o}$, quite close to the spherical light-front. The circle represented by points $P$ on the spherical light-front $r=ct$ where the field strength is maximum as a function of $\theta$, lies almost vertically above $z_{\rm o}$, the `present' position of the charge, and thus are not very far from it, implying that the field at large $r$ is still around the `present' location of the charge.}
\end{center}
\end{figure*}
%----------------------------------------------------------

Let the charge moving with a uniform acceleration, $a\equiv \gamma ^{3} \dot{v}$ along $+z$ axis, was momentarily stationary at time $t=0$ at a point $z=\alpha$, chosen, without any loss of generality, so that $\alpha=c^{2}/a$.
The position and velocity of the charge, before or after, at any other time $t$ are then given by \cite{5,10,88}
$z_{\rm o}=(\alpha^{2}+c^{2}t^{2})^{1/2}$, $v_{\rm o}=c^{2}t/z_{\rm o}$ and $\gamma_{\rm o}=z_{\rm o}/\alpha$, which implies $ct=\alpha\gamma_{\rm o}v_{\rm o}/c$.

In a typical radiation scenario, the radiated energy moves away ($r\rightarrow \infty$), with the charge responsible remaining behind, perhaps not very far from its location at the corresponding retarded time, e.g., in localized charge or current distributions in a radiating antenna. This of course necessarily implies that not only the motion of the charge is bound, its velocity and acceleration are having, some sort of oscillatory behaviour, even if not completely regular. However, in the case of a uniform acceleration, such is not the case.
Due to a constant acceleration, the charge picks up speed, and after a long time its motion will become relativistic, with $v \rightarrow c$ and the corresponding Lorentz factor becoming very large ($\gamma\gg 1$). Then, due to the relativistic beaming, the distant fields of the charge as well as the associated Poynting flux is appreciable only within a narrow cone-opening angle, with a maxima at $\theta \sim 1/\gamma$ \cite{1,2,25}, about the direction of motion. 

One comes across such instances of relativistic beaming in the synchrotron radiation, where due to an extremely relativistic motion ($v \approx c$) of the gyrating charge, the radiation is confined to a narrow angle $\sim 1/\gamma$ about the instantaneous direction of motion \cite{1,RG70}. Furthermore, in extragalactic radio sources, due to highly relativistic motion of a radio source component with respect to the observer's frame of reference, the radio emission appears confined to a narrow cone of emission with a cone-opening angle  $\sim 1/\gamma$ \cite{RE66}.

In our present case, the charge, moving with a velocity $v \rightarrow c$, is not very far behind the spherical light-front of radius $r=ct$. The charge, with $v\approx c(1-1/2\gamma^2)$, moves a distance $\sim ct(1-1/2\gamma^2)$ along the $z$-axis, while the circle of maxima of the field, represented by $P$ at $r=ct$, has moved along the $z$-axis a distance, $r\cos\theta\approx ct(1-\theta^2/2)\approx ct(1-1/2\gamma^2)$, thus the field maxima lies in a plane normal to the $z$-axis that passes nearly through the `present' position of the charge on the $z$-axis (Fig.~2), and the fields are all around the charge. 
The electric field, in fact, very much resembles that of a charge moving with a uniform velocity equal to the `present' velocity of the uniformly accelerated charge, with the field in a plane normal to the direction of motion. Thus, as the fields move toward infinity, so does the charge and the fields are confined along the direction of motion in a small, finite region  $ct/2\gamma^2\sim \alpha/2\gamma$ about the `present' position of the charge, very similar to the  uniform velocity case where electric field is ever appreciable only near the `present' position of the charge, in a region whose extent falls as $1/\gamma$ . As was shown in Section 3.1, the fields actually are the self-fields of the charge that due to the acceleration fields, increase in strength, as the charge picks up speed, to a value expected from that of the charge moving with a uniform velocity equal to the `present' velocity of uniformly accelerated charge, and accordingly, there is no radiation being `emitted' by the charge. 

%---------------------------------------------------------
We can verify the above statements explicitly by a comparison of the fields of a uniformly accelerated charge, which may have a relativistic `present' velocity $v_{\rm o}\rightarrow c$ and a corresponding Lorentz factor $\gamma_{\rm o}\gg 1$, with those of a charge moving with a uniform motion, with exactly the same velocity $v_{\rm o}$ and thus the same Lorentz factor $\gamma_{\rm o}$. 

The electromagnetic fields of the charge moving with a uniform acceleration, is given in cylindrical coordinates ($z,\rho,\phi$), as \cite{5,10,88} 
\begin{eqnarray}
\label{eq:89.4s}
E_{z}&=&-4e\alpha^{2}(\alpha^{2}+c^{2}t^{2}+\rho^{2}-z^{2})/\xi^{3}\nonumber\\
E_{\rho}&=&8e\alpha^{2}\rho z/\xi^{3}\nonumber\\
B_{\phi}&=&8e\alpha^{2}\rho ct/\xi^{3}\;,
\end{eqnarray}
where $\xi=[(\alpha^{2}+c^{2}t^{2}-\rho^{2}-z^{2})^{2}+4\alpha^{2}\rho^{2}]^{1/2}$. 

The above solution is restricted to a region $z>-ct$ with a discontinuity in the fields at $z=-ct$ \cite{5,10,88}. 
These field expressions are equivalent to the field expressions in terms of retarded-time quantities, and can be derived in the case of a uniformly accelerated charge starting from  Eq.~(\ref{eq:1a}), using algebraic transformations \cite{88}.
%-------------------------------------------------------------------

On the other hand, the electromagnetic field of the charge moving with a uniform velocity $\bf v_{\rm o}$, can be written in a spherical coordinates ($R,\Theta$), or in cylindrical coordinates ($\rho,\Delta z$), 
centered at the ``present'' charge position \cite{1,2,25}, as
\begin{eqnarray}
\label{eq:89.4t}
{\bf E}&=&\frac{e\hat{\bf R}}{R^{2}\gamma_0^{2}[1-(v_0/c)^{2}\sin^{2}\Theta]^{3/2}}=\frac{e\gamma_{\rm 0}{\bf R}}{[\rho^2+\gamma_{\rm 0}^2\Delta z^{2}]^{3/2}}\,,
%E_{R} & = & \frac{e}{R^{2}\gamma_{\rm o}^{2}(1-(v_{\rm o}/c)^{2}\sin^{2}\theta)^{3/2}}\;.
\end{eqnarray}
The magnetic field in both cases is given by $\bf {B= v_{\rm o}\times E}$. Equation (\ref{eq:89.4t}) can be derived in the case of a uniformly moving charge from velocity fields (the first term in the square brackets in Eq.~(\ref{eq:1a})) \cite{1,2,25}.

As is well known, for a charge moving relativistically with a uniform velocity, the electric field component perpendicular to the direction of motion is stronger by a factor $\gamma$ relative to the component along  the direction of motion, with the field lines becoming oriented perpendicular to the direction of motion \cite{1,2,25}. Moreover, for a large $\gamma$, the field becomes negligible, except in a narrow zone along the direction of motion, with the field lines confined mostly within a small angle, $\Delta z/\rho \sim 1/\gamma$, with respect to a plane normal to the direction of motion and passing through the `present' charge position \cite{88}. 
%----------------------------------------------------------
\begin{figure}[t]
\begin{center}
\includegraphics[width=\linewidth]{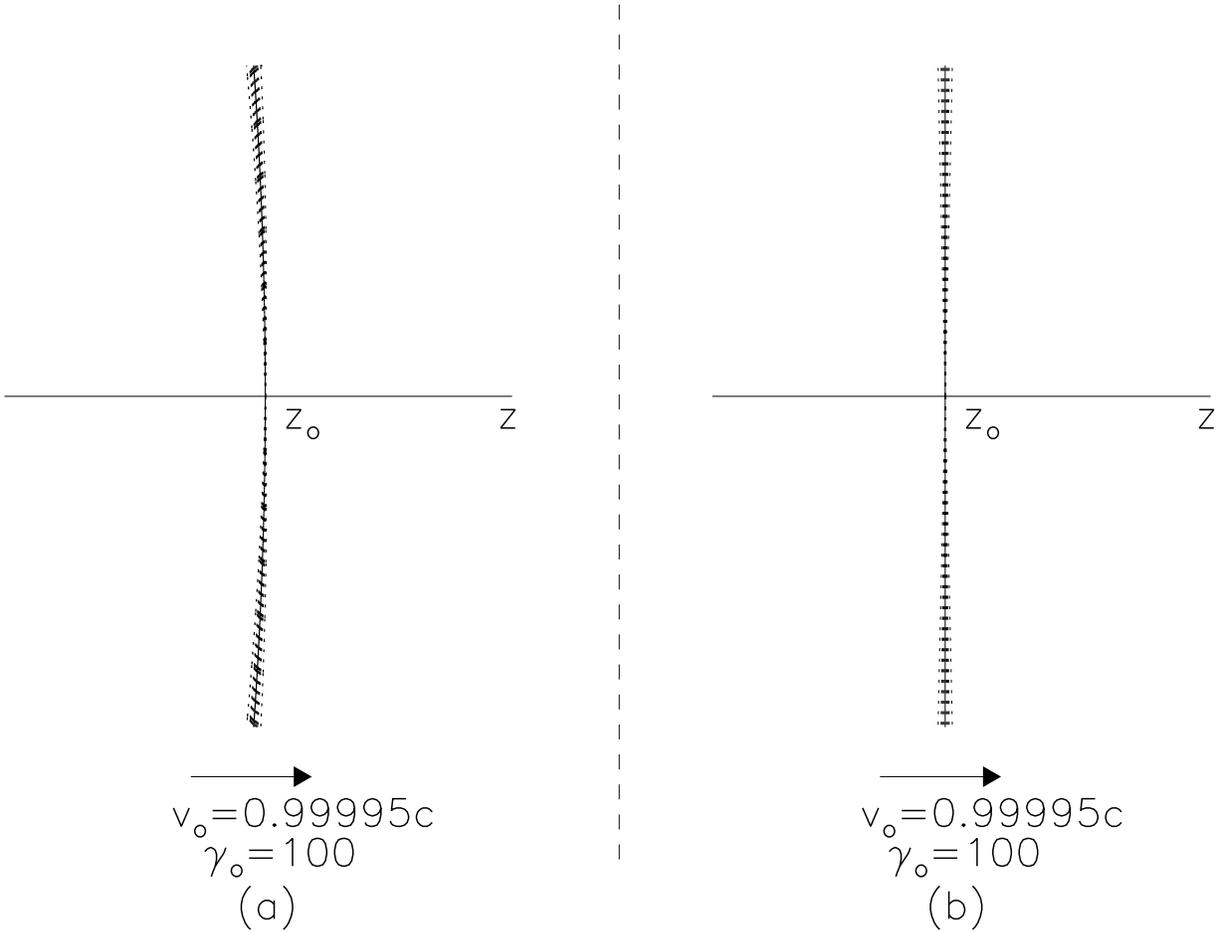}
\caption{The electric field distribution (a) of a uniformly accelerated charge, with a `present' velocity $v_{\rm o}=0.99995 c$, corresponding to $\gamma_{\rm o}=100$ (b) of a charge moving with a uniform velocity $v_{\rm o}=0.99995 c$, corresponding to $\gamma_{\rm o}=100$. In both cases, the electric field lines are confined mostly within a small angle $\sim 1/\gamma$ with respect to the electric field lines that begin from the charge position $z_{\rm o}$, in  plane perpendicular to the direction of motion.}
\end{center}
\end{figure}
%----------------------------------------------------------
%----------------------------------------------------------
\begin{figure}[t]
\begin{center}
\includegraphics[width=11cm]{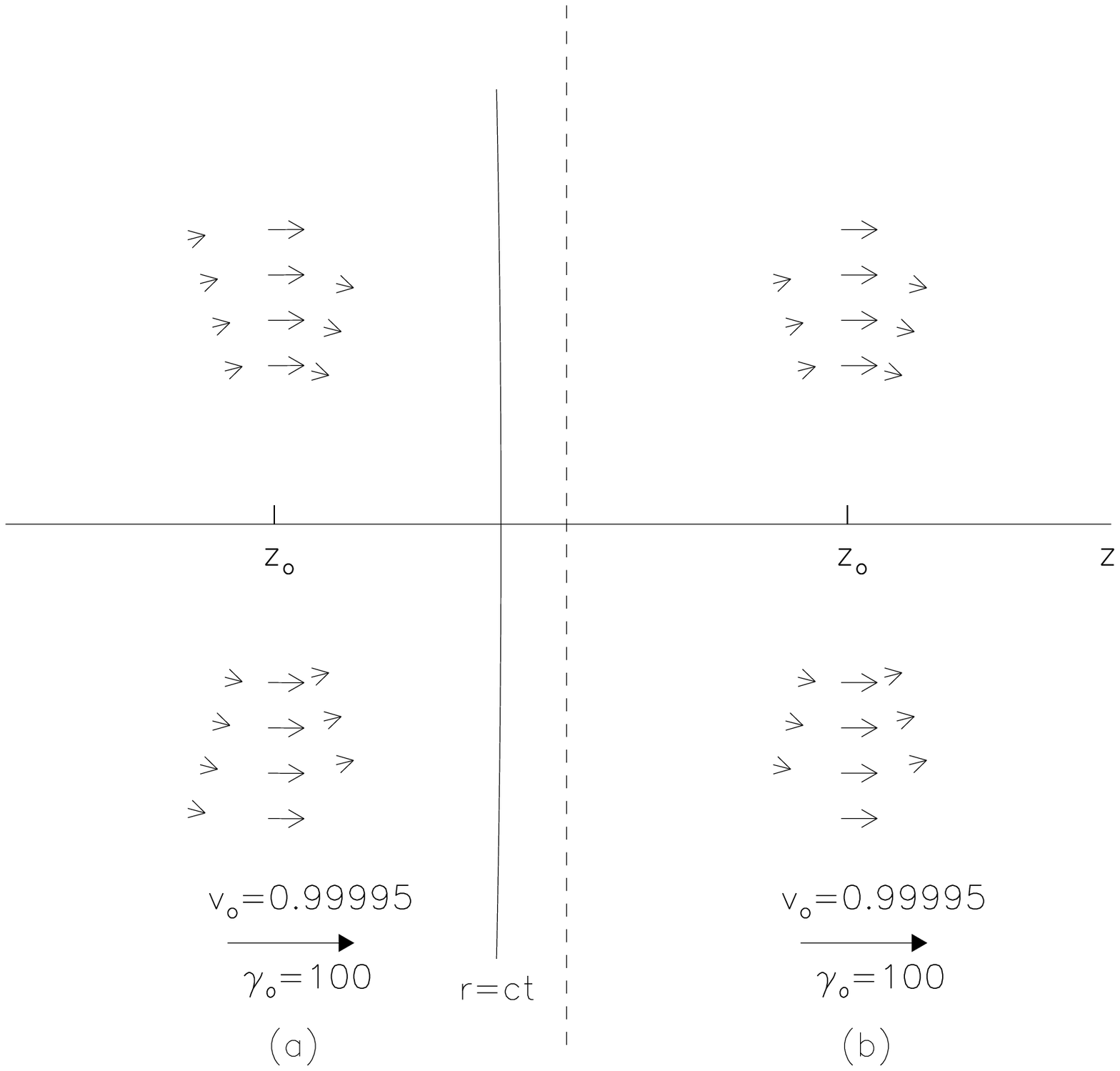}
\caption{The Poynting vector for a charge (a) moving with a uniform proper acceleration, and is presently at $z_{\rm o}$ moving with a `present' velocity $v_{\rm o}=0.99995c$, corresponding to $\gamma_{\rm o}=100$ (b) moving with a uniform velocity $v_{\rm o}=0.99995c$, corresponding to $\gamma_{\rm o}=100$. The spherical light-front $r=ct$ is shown in the case of uniformly accelerated charge, which looks like a plane on this scale. The overall Poynting flow in both cases is along the direction of motion of the charge. Arrows show Poynting vector directions at different distances from the charge. The length of an arrow is not a direct indicator of the magnitude of the corresponding Poynting vector, the plot shows the trend only qualitatively. In fact, the magnitude of the Poynting vector, represented by larger arrows, is maximum at the plane normal to the direction of motion, passing through the charge at $z_{\rm o}$,  and drops  rapidly off the plane. At the positions of smaller arrows, shown in the figure, the magnitude of the Poynting vector falls as much as by a factor of  $\sim 10^8$.}
\end{center}
\end{figure}
%----------------------------------------------------------

Now if we plot the electric field (Eq.~(\ref{eq:89.4s})) of the uniformly accelerated charge, for a large $r=ct$, which also implies $v_{\rm o} \rightarrow c$ and $\gamma_{\rm o}\gg 1$, and compare it with the field (Eq.~(\ref{eq:89.4t})) of a charge  moving with the same, but a uniform, velocity $\bf v_{\rm o}$ and thus having the same $\gamma_{\rm o}$, we find that the fields are quite similar in both cases. Figure 3 shows a comparison of the electric fields in both cases for $\gamma_{\rm o}=100$, corresponding to  $v_{\rm o}=0.99995c$. In both cases fields are very similar and extend, from the ``present'' charge position, in direction normal to the direction of motion. 

Figure 4 shows the corresponding Poynting flow, almost indistinguishable in both cases,  with the overall Poynting flow in each case being along the direction of motion of the charge, confirming that the Poynting flow for a uniformly accelerated charge merely represents the ``convective'' flow of self-fields, along with the moving charge, like in the case of a charge moving with a uniform velocity. Of course, in the case of a uniformly accelerated charge, the self-field strength continuously keeps getting `updated' due to acceleration fields, in tune with the changing charge velocity due to its uniform acceleration.
Naturally, there is no radiation reaction in the case of a uniformly accelerated charge since no field energy is being `radiated away' from such a charge. This, of course, also makes the case of a uniformly accelerated charge fully conversant with the strong principle of equivalence.

In order to avoid a contradiction with Larmor's radiation formula, it has been suggested that the radiation emitted from the uniformly accelerated charge goes beyond the horizon, in regions of space-time inaccessible to an observer co-accelerating with charge \cite{10,AL06}. Actually, it is a misconception as from Eq.~(\ref{eq:89.4s}), $E_{\rho}=0$ at the $z=0$ plane for all $t$, implying that there is no component of Poynting flux through the $z=0$ plane ever. This statement is true for all inertial frames at all times; the only exception is at $t=0$ when an infinite $z$-component of Poynting vector due to  $\delta$-fields is present at $z=0$,  
causally related to the charge during its uniform velocity before an acceleration was imposed at an infinite past. The $\delta$-field, is, in fact, not causally related to the charge during its uniform acceleration, whose influence at that time lies only in the $z>0$ region. All fields, originating from the accelerating charge positions, lie in the region $z>0$ at time $t=0$ and the radiation, if any, from the accelerating charge should also be present there only and not appear at the horizon at $z=0$. 
In fact, it has been shown that because of a rate of change of
acceleration at the time when the acceleration was first imposed on the charge, an event with which the $\delta$-field has a causal relation, the charge underwent radiation losses \cite{68a}, owing to the Abraham-Lorentz radiation reaction \cite{abr05,16,20,68b}, thereby neatly explaining the total energy lost by the charge into $\delta$-field during a transition from a uniform velocity phase to the uniform acceleration phase at infinite past \cite{88}.
In fact, as has been demonstrated here, all fields, including the acceleration fields, having a genesis from the uniform accelerated charge, remain around the moving charge and are not {radiated away} or  dissociated from the charge as long as it continues moving with a uniform acceleration. 

%---------------------------------------------------------------
\section{Conclusions}
We showed that a discrepancy between two formulations of the power going into electromagnetic radiation and the mechanical power loss of the radiating charge, merely reflects the difference in the power going in self-fields of the charge between the retarded and  present times. It was shown that equating the Poynting flux at time $t$, given by Larmor's formula, to the mechanical
 power loss of the charge at a retarded time $t'$,  is not in accordance with Poynting's theorem, where all quantities need to be calculated, strictly for the  same instant of time. It was further shown that in the case of a periodic motion, where there is no difference in the radiated power averaged over the period starting at $t$ or  $t'$, Larmor's formula does yield a correct, average power loss by the charge, an argument which, however, cannot be applied in the case of a uniformly accelerated charge. It was shown that for a charge uniformly accelerated, all its fields, including the acceleration fields, originating from the time retarded positions of the charge, are not {\em radiated away} from it and remain around the `present position of the charge as its self-fields. 

%%%%%%%%%%%%%%%%%%%%%%%%%%%%%%%%%%%%%%%%%%
\funding{This research received no external funding.}
%%%%%%%%%%%%%%%%%%%%%%%%%%%%%%%%%%%%%%%%%%
%\acknowledgments{In this section you can acknowledge any support given which is not covered by the author contribution or funding sections. This may include administrative and technical support, or donations in kind (e.g., materials used for experiments).}

%%%%%%%%%%%%%%%%%%%%%%%%%%%%%%%%%%%%%%%%%%
\conflictsofinterest{The author declares no conflict of interest.} 
%%%%%%%%%%%%%%%%%%%%%%%%%%%%%%%%%%%%%%%%%%
\reftitle{References}
{}
\end{document}